\documentclass{IEEEtran}
\usepackage{cite}
\usepackage{amsmath,amssymb,amsfonts}
\usepackage{graphicx,color}
\usepackage{textcomp}
\usepackage{epstopdf}

\usepackage{epsfig}
\usepackage{amsmath}
\usepackage{amsthm}
\usepackage{amsfonts}
\usepackage{amssymb}
\usepackage{hyperref}
\hypersetup{colorlinks,%
citecolor=black,%
filecolor=black,%
linkcolor=black,%
urlcolor=black,%
pdftex} 
\usepackage{graphicx}
\usepackage{graphics}
\usepackage{float}
\usepackage{url}
\usepackage{algorithm}
\usepackage{algpseudocode}
\usepackage{setspace}
\usepackage{subfigure}
\usepackage{balance}
\usepackage{amsmath} 
\usepackage{bm}      

\usepackage{hyperref}
\usepackage{color}

\usepackage{hyperref}
\usepackage[normalem]{ulem}
\useunder{\uline}{\ulined}{}%
\hypersetup{
    colorlinks=blue,
    linkcolor=blue,
    filecolor=blue,      
    urlcolor=blue,
}

\def\mi0{\mathbf{0}}
\def\mih{\mathbf{h}}

\def\mix{\mathbf{x}}

\def\BibTeX{{\rm B\kern-.05em{\sc i\kern-.025em b}\kern-.08em
    T\kern-.1667em\lower.7ex\hbox{E}\kern-.125emX}}
\AtBeginDocument{\definecolor{ojcolor}{cmyk}{0.93,0.59,0.15,0.02}}

\begin{document}

\title{BIA Transmission in Rate Splitting-based Optical Wireless Networks}

\author{\IEEEauthorblockN{Ahmad Adnan Qidan, Khulood Alazwary, Taisir El-Gorashi, Majid Safari, Harald Haas,  Richard V. Penty,  Ian H. White, and Jaafar M. H. Elmirghani}\\

}
\maketitle

\begin{abstract}
Optical wireless communication (OWC) has recently received massive interest as a new technology that can support the enormous data traffic increasing on daily basis. In particular, laser-based OWC networks can provide terabits per second (Tbps) aggregate data rates. However, the emerging OWC networks require a high number of optical access points (APs), each AP corresponding to an optical cell,  to provide uniform coverage for multiple users. Therefore, inter-cell interference (ICI) and multi-user interference (MUI) are crucial issues that must be managed efficiently to provide high spectral efficiency. In radio frequency (RF) networks,  rate splitting (RS) is proposed as a transmission scheme to serve multiple users simultaneously following a certain strategy. 
It was shown that RS provides high data rates compared to orthogonal and non-orthogonal interference management schemes. Considering the high density of OWC networks, the application of RS within each optical cell might not be practical due to severe ICI. 
In this paper, a new strategy is derived referred to as blind interference alignment-rate splitting (BIA-RS) to fully coordinate the transmission among the optical APs, while determining the precoding matrices of multiple groups of users formed beforehand. Therefore, RS can be implemented within each group to manage MUI. The proposed BIA-RS scheme requires two layers of power allocation to achieve high performance. Given that, a  max-min fractional optimization problem is formulated to optimally distribute the power budget among the groups and the messages intended to the users of each group. Finally, a power allocation algorithm is designed with multiple Lagrangian multipliers to provide practical and sub-optimal solutions.  The results show the high performance of the proposed scheme compared to other counterpart schemes.     
\end{abstract}

\IEEEpeerreviewmaketitle

\maketitle
\section{Introduction}
\IEEEPARstart{T}he use of the Internet has massively increased in recent years due to  emerging technologies  such as Internet of Things (IoT), robotics, video streaming, 3D printing and virtual reality (VR), etc. This unprecedented growth in data traffic causes several challenges on current  radio frequency (RF) networks including lack of resources,  power consumption and secrecy. Therefore, researchers  in both industrial and academic communities  have investigated  the possibility to define new technologies  that can work  with RF networks to support high-user demands and relax traffic-congestion in the next generation (6G) of  wireless communications \cite{8240590}. Optical wireless communication (OWC) has received  attention as a promising technology with the potential  to overcome the drawbacks of RF networks where the optical band offers license-free bandwidth, improved secrecy and usually optical access points (APs) provide high energy efficiency compared with RF APs. An OWC network can provide an aggregate data rate in a range  of gigabit per second (Gbps) using conventional light emitting diodes (LEDs) \cite{8240590,hhja}. Like other technologies, OWC faces  several challenges   such as  light blockage, the confined  converge area of the optical AP, and the low modulation speed of LED light sources \cite{7239528,6685754}. It is worth mentioning that increasing the number of LED-based optical APs in an indoor environment can  provide a wide coverage area and seamless user-transition. However, LEDs usually are installed primarily for illumination, and therefore, increasing the number of LEDs is limited by the recommended illumination levels in such indoor environments. Alternatively, Infrared lasers such as  vertical-cavity surface-emitting lasers  (VCSELs) can be used for data transmission  since they have high modulation speeds compared to LEDs, and they are commercially available at low prices. In an indoor environment, clusters of VCSELs can be deployed on the ceiling  to provide uniform coverage. However, the total radiated power of the VCSELs must be within eye safety regulations. It was shown that VCSEL-based OWC networks can provide up to terabits per second (Tbps) aggregate data rates \cite{mo6963803,AA19901111,owc-sp22}.

Interference management is a crucial issue in high density multiple-input and multiple-output 
(MIMO) OWC scenarios where multiple users must be served simultaneously to enhance spectral efficiency. Various orthogonal techniques 
can be applied to allocate resources in an orthogonal fashion among users. Despite the avoidance of interference, users might experience low quality of service due to lack of resources.  Recently,  non-orthogonal multiple access (NOMA) \cite{7342274,7572968} has been proposed for OWC to manage multi-user interference (MUI) through exploiting the power domain at the transmitters where users are classified as strong and weak users, and then, power allocation is performed considering the channel quality of each user. It was shown that NOMA provides  high spectral efficiency in  OWC  compared with other orthogonal transmission schemes due to the fact that  the resources of the network can be reused among users at a given time \cite{7572968}. However, if  users have comparable   channel gains, the application of NOMA becomes a challenge. Besides, the strong users in NOMA might perform multi-layer successive interference cancellation (SIC) to decoded the information intended to all other users, which results in high complexity for such users.    

\subsection{Related Works}
For RF networks, a new transmission scheme referred to as rate splitting (RS)  was proposed in \cite{7470942} to decode part of the interference, while treating the other part as noise.  RS is usually defined  as a bridge between  two different  interference management strategies, orthogonal and non-orthogonal. Basically, in a wireless network serving multiple users, the methodology of RS relies on splitting  the message of each user into  common and private messages in order to combine all the common messages of the users  into a super common message superimposed on top of  the private messages.  In RS, users first decode the super common message with minimum error probability, and remove it from the received signal using SIC. Then, each user decodes  its private message while treating  the other private messages as noise. Interestingly, the power must be  carefully allocated between the common and private messages to minimize MUI and maximize the spectral efficiency of the network.  
In \cite{6289367}, a closed-form strategy for RS  was  proposed  to serve $ K $ users in a time correlated multiple-input single-output (MISO) broadcast channel scenario. It was shown that the strategy of splitting the message of each user into two parts offers higher sum rates compared to traditional interference management schemes  schemes such  as zero forcing (ZF) and TDMA.  In \cite{7513415}, the RS strategy was applied  in a downlink multi-user MISO system to achieve high quality of service under low power consumption. In \cite{7434643}, a novel scheme called  hierarchical rate splitting (HRS) using two-layers of RS was proposed to manage intra-group and inter-group interference in MIMO networks. 
In \cite{7805217},  a topological
RS (TRS) scheme  with weighted sum interpretation is derived in $ L $-cell MISO interference channel scenarios.
 In  HRS and TRS, users are arranged into groups, and multiple common messages are transmitted to  the formed groups where each common message contains the private messages of the users belonging to a certain group. The superiority of HRS and TRS was demonstrated compared to traditional RS in terms of the data rate achieved  due to the fact that  theses more advanced schemes reduce the noise resulting  from the use of SIC at each user.

In the context of OWC, the application of RS has been considered in \cite{Naser2020,HRS,Kh9500371}. Specifically, in \cite{Naser2020}, RS was used to manage  the interference between two users served by a signal optical cell deployed in an indoor environment where an optimization problem is formulated to maximize the sum rate under the power constraint of the optical transmitter.  In \cite{HRS}, the performance of RS and HRS schemes  were evaluated in a multi-user signal cell OWC network, where it was shown that the data rate achieved by RS and/or HRS  decreases considerably as the number of users increases due to noise enhancement.  In \cite{Kh9500371}, an optimization problem was formulated to allocate the power of the optical cell among the messages of the RS scheme with the aim of enhancing the sum rate of the network. However, the application of the proposed RS schemes in the literature is not straightforward  when it is considered for high density OWC networks. In other words, the emerging OWC networks using lasers are composed of a high number of optical APs  due to the confined coverage area of these optical APs. Therefore, users might experience severe inter-cell interference (ICI) if the transmission among the optical APs is not coordinated properly. In \cite{9145287}, the RS scheme was applied in a multiple cell OWC to manage MUI within the coverage of each cell, while a coordinated multi-point transmission (CoMP) technique was implemented to enhance signal to noise ratio (SNR) for users located at the edges of the cells. It is worth mentioning that CoMP is enabled in an OWC network through  backbone links connecting all the optical APs together to share the data streams intended to  all the users \cite{6825144}. 
One of the limitations in using CoMP for ICI  in laser-based OWC networks is that users located at the center of their corresponding cell  are  also subject to ICI from the adjacent cells due to the small coverage area of each optical AP. One of the solutions for this issue is to consider a frequency reuse (FR) approach or to implement ZF on top of CoMP to cancel ICI at the users located at the center of each cell. Note that,   the implementation of  FR might affect the spectral  efficiency of the network considerably. In addition, ZF is known by its low performance at low SNR regime, and the design of the ZF precoder along with CoMP involves high complexity especially in RS-based OWC networks where power allocation is essential to address \cite{9693949}. 

Recently,  blind interference alignment (BIA) has been considered as a promising transmission scheme due to its unique interference management strategy that offers coordination among multiple optical APs and allocate non-orthogonal resources to all users with the guarantee of MUI management \cite{GWJ11}. Interestingly, applying  BIA can be possible if the users receive linearly independent channel responses from all the transmitters. In \cite{MPGV18}, a reconfigurable optical detector composed of multiple photodiodes is derived to apply BIA in OWC networks.  In \cite{8636954}, it was shown that BIA is more suitable for OWC networks compared to ZF and  orthogonal transmission schemes  due to its ability to fully coordinate the transmission among the transmitters, turning the ICI signals into useful signals for all the users regardless of their locations. However, BIA suffers performance degradation as the number of users  increases due to the requirements of the channel coherence time and noise enhancement \cite{8636954,9064520,9500371}.
It is worth mentioning that, the strategy of BIA can be redesigned as a precoder that achieves full coordination among the optical APs, and manages the interference among multiple groups of users rather than MUI in order to relax the BIA limitations \cite{8935164}.

The discussion above has motivated this work to design a robust transmission scheme referred to as BIA-RS from now on, which combines the features of BIA and RS, where the BIA precoder manages ICI and the interference among multiple groups of users, while RS manages MUI. In the literature of BIA, uniform power allocation schemes are usually considered among the resources allocated to multiple users. However, the proposed BIA-RS needs a new power allocation scheme that guarantee a high quality of service for multiple users. In the following, the  main contributions of this work are  reported in detail.  

\subsection{Main contributions}
We consider an OWC network composed of multiple APs using VCSEL, which is a strong candidate in the emerging 6G OWC networks. Each  optical AP operates under eye safety regulations, and illuminates a confined area. All the optical APs overlap among each other causing ICI for multiple users distributed on the communication floor. In this work, the users are divided into multiple groups to relax noise enhancement, and the ICI signals are managed as  useful signals using a precoder designed according to the BIA strategy that also manages inter-group interference, and therefore, RS can be  implemented successfully within each group to manage MUI. In contrast to the literature, the integration of  BIA and RS is investigated for the first time in this paper. Additionally, the strategy of BIA simply assumes a fixed power allocation approach, which can cause high power consumption for the proposed BIA-RS scheme. Therefore, power allocation is investigated to enhance the overall performance of the network and guarantee high quality of service for the users. The main contributions can be listed as follows 
\begin{itemize}

\item The proposed BIA-RS scheme is derived for interference management in a laser-based OWC network, which involves two steps. First, BIA  operates as an outer precoder where a transmission block comprising  resource blocks is built from the whole set of optical APs, i.e., full coordination. These resource blocks are  determined according to the number of optical APs and groups of users, and they are allocated in a non-orthogonal fashion with the guarantee of inter-group interference management. Second, RS is applied for MUI within each group where the power allocated to a certain group is divided  between two different messages, common and private messages.  Furthermore, the achievable sum rate of the proposed BIA-RS scheme  is derived to define it as a general framework applicable to different scenarios.

\item Power allocation is crucial in OWC networks where the power budget must be used efficiently. In the context of BIA-RS, two layers of power allocation, i.e., power allocation among the groups of users and power allocation among the messages intended to the users of each group, are needed to maximize the overall sum rate of the network. Interestingly, a fixed power allocation approach among the groups can be considered, as well as among the common and private messages. However, it might result in waste of power  due to the fact that users belonging to a certain group might not use their power fully. In this context, an optimization problem is formulated for power allocation under certain constraints to maximize the minimum sum rate within each group. 

\item The formulated optimization problem is defined as  max-min fractional program with a particular structure, which is difficult to solve, Therefore, the optimization problem is reformulated to solve using the parametric approach. The new objective function contains four Lagrangian multipliers. Two of these multipliers guarantee low power consumption, while the other two multipliers guarantee that the users belonging to each group experience high quality of service. The overall power allocation algorithm  iterates over these multipliers until a significant sub-optimal solution is reached and the sum rate of the network is maximized. 
\end{itemize}
The results demonstrate that our proposed BIA-RS scheme achieves high performance in terms of sum data rate, BER and energy efficiency in dense OWC networks compared to BIA, NOMA, RS and two baseline schemes derived according to the majority of work in the literature.

The remainder of this paper is organized as follows. In Section \ref{sec:system}, the system model of the OWC network considered is described.  In Section \ref{sec:CSIRS}, the proposed BIA-RS scheme is derived in detail.  The formulation and analysis of power allocation for the proposed scheme are derived in Section \ref{sec:power}. Finally, Section \ref{sec:re} presents simulation results, and Section \ref{sec:con} provides concluding remarks and future work. 
 
{\it Notation}.  The notations considered in this work are defined as follows. First,  matrices and vectors are denoted by the bold upper case and lower case letters,  respectively. To represent identity and zero matrices with $M\times M$ dimension, we call out to   $\mathbf{I}_M$ and $\mathbf{0}_M$ notations, respectively, while $\mathbf{0}_{M,N}$  denotes a $M\times N$ zero matrix, $[\,\,]^T$ and $[\,\,]^H$ indicate  the transpose and hermitian transpose operators, respectively. Finally, $\mathrm{col}\{\}$ is the column operator that stacks the considered vectors in a column.
\begin{figure}[t]
\begin{center}\hspace*{0cm}
\includegraphics[width=1\linewidth]{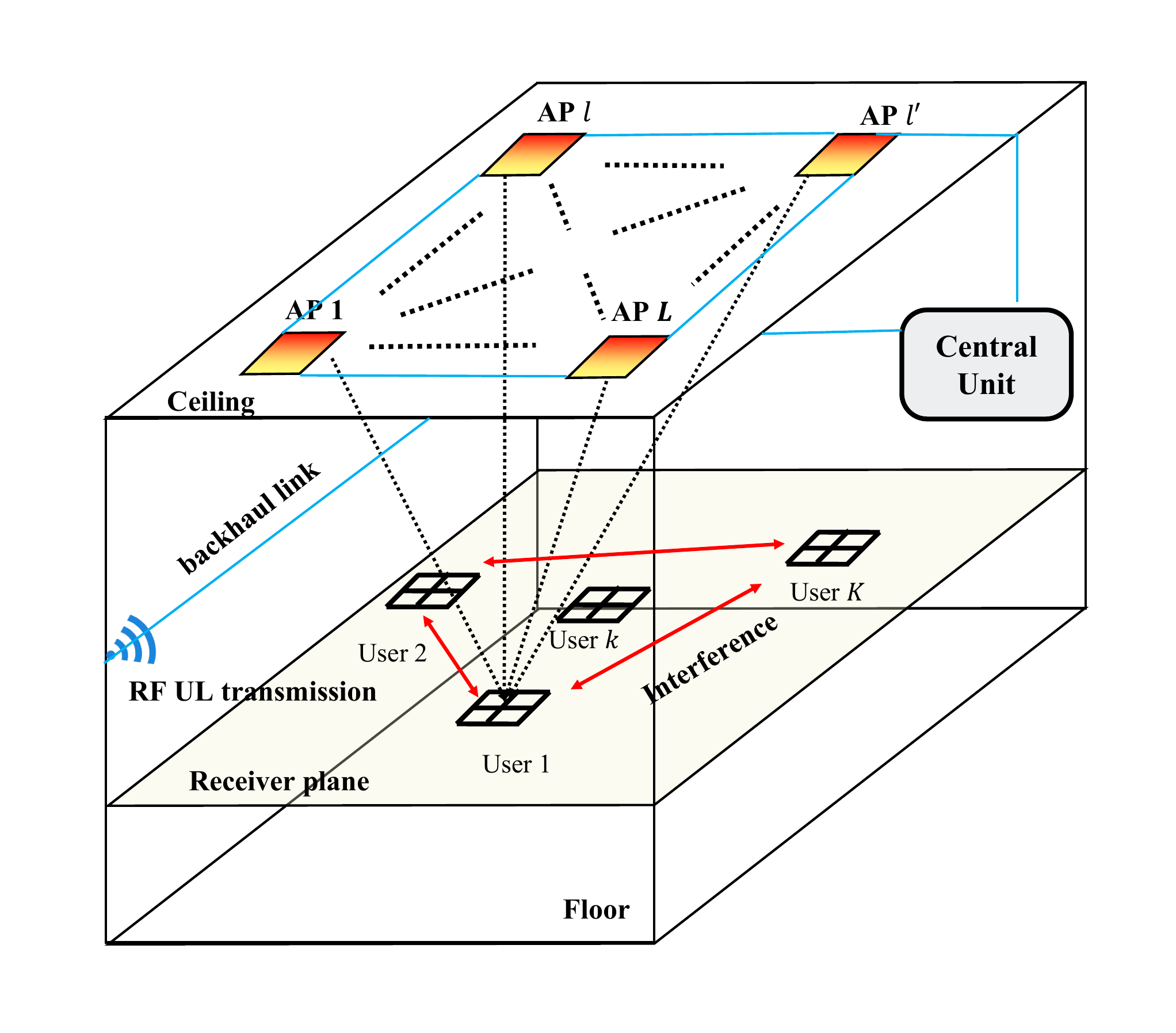}
\end{center}
\vspace{-2mm}
\caption{An OWC network composed of a number of optical APs serving multiple users.}\label{Figmodel}
\vspace{-5mm}
\end{figure}
\section{System Model}
\label{sec:system}
An indoor downlink OWC network  is considered as  shown in Fig.~\ref{Figmodel}.{  We define a set, $\mathcal{L}$, consisting multiple optical APs, $ L, l= \big\{1, \dots, L \big\} $, installed on the ceiling where each optical AP is an array of $ L_{v} \times L_{v} $ VCSELs in order to expand its coverage area. On the receiving plane, a set, $\mathcal{K}$, is defined consisting multiple users, $ K, k= \big\{1, \dots, K \big\} $, distributed randomly where each user is equipped   with an optical detector with $ M, m= \big\{1, \dots, M \big\} $, photodiodes to guarantee wide field of view (FoV) \cite{MPGV18,8636954}. The users can be arranged into a set, $\mathcal{G}$, consisting multiple groups, $ G, g= \big\{1, \dots, G \big\} $, where each group, $g $, is formed of a set, $\mathcal{K}_g$, consisting multiple users, $ K_{g}, k= \big\{1, \dots, K_{g} \big\} $.} The groups of users are formed using a well known clustering algorithm called $ K $-means algorithm \cite{8636954} where users located at a distance less than or equal to a threshold distance $ d_{th} $ are more likely to form a group. In the network considered, users experience severe ICI, which is managed in a certain way as explained in  the rest of the paper where the transmission among $ L $ optical APs is coordinated forming a large-scale optical cell. Given that, the received signal at  photodiode $ m $ of a generic user $ k $, i.e., without considering the formation of $ G $ groups,  from $ L $  optical APs   can be written as
\begin{equation}
y^{[k]}[n]=\zeta\mih^{[k]}(m^{[k]}[n])^{T} \mix[n]  + z^{[k]}[n],
\end{equation}
{where $\zeta$ is the receiver responsivity}, $ \mih^{[k]}(m^{[k]}[n])^{T} \in \mathbb{R}_+^{L\times 1} $, $ m^{[k]}[n] $  is the channel state  of photodiode $ m $ at time slot $ n $, and  $ \mix= \rho  \sqrt{P^{[k]}} \mathbf{s}^{[k]}+ \rho\sum^{K}_{k'\neq k} \sqrt{P^{[k']}} \mathbf{s}^{[k']}+ \rho I_{Dc} \times \mathbf{1}_{L\times1} $ is the transmitted signal where $ \rho $ is the electric to optical power conversion factor,  $ P^{[k]} $ is the power allocated to user $ k $, $ \mathbf{s}^{[k]}$ is the useful information of user $ k $, $ \sum^{K}_{k'\neq k} \sqrt{P^{[k']}} \mathbf{s}^{[k']} $  is the information sent to other users, and $ I_{Dc}  $ is the direct current that guarantees the non-negativity of the transmitted signal {\footnote{ We consider a fixed value for $ I_{Dc} $  ensuring that the VCSEL transmitters work in the linear
region. Note that, enhancing the DC biasing current can be achieved through complex optimization problems, which is not the scope of this work.}}, {which can be removed from the received signal using a capacitor}. Moreover, $ z^{[k]} $ is real valued additive white Gaussian noise with zero mean and variance determined as contributions from  shot noise,  thermal noise and the intensity noise of the laser \cite{AA19901111}. In this work, all the optical APs are  connected through a central unit (CU) that has information regarding the distribution of the users and the channel coherence time. It also controls the resources of the network such that optimization problems with different contexts can be performed to enhance the performance of the network. Moreover, a WiFi AP is deployed to provide uplink transmission where  users can forward  the information needed for solving the optimization problems.


\subsection{Transmitter and receiver}
We consider each optical AP as an array of VCSELs  operating under eye safety regulations \cite{mo6963803, AA19901111}. Basically, the VCSEL transmitter has Gaussian beam profile dictated by the DC bias current applied, and its 
 power distribution is determined based  on the beam waist $ W_{0} $, the wavelength $ \lambda $ and the distance $ d $ through which the beam travels.  In this context,  the beam radius of the VCSEL  at the receiving plane located at distance $ d $  is given by 
\begin{equation}
W({d})=W_{0} \left( 1+ \left(\frac{d}{d_{Ra}}\right)^{2}\right)^{1/2},
\end{equation}
where $ d_{Ra} $ is the Rayleigh range determined by $ d_{Ra}= \frac{\pi W^{2}_{0} n }{ \lambda},
$ where $ n $ is the refractive index of the medium, which is  air in this case, i.e., $ n=1 $. Moreover, the spatial distribution of the intensity of the VCSEL over the transverse plane at distance $ d $ can be expressed as  
\begin{equation}
I_{t}(r,{d}) = \frac{2 P_{tr}}{\pi W^{2}({d})}~ \mathrm{exp}\left(-\frac{2 r^{2}}{W^{2}({d})}\right),
\end{equation}
where $ r $ is the radial distance from  the beam spot center and  $ P_{tr} $ is the transmit power per the beam. On the user side, each optical detector  is composed of $ M $ photodiodes{\footnote{Usually, the number of photodiodes in such optical detectors must be at least equal to the number of the optical APs in the environment to guarantee the ability to provide linearly independent channel responses \cite{mo6963803, AA19901111}. }}
with a whole detection area given by  $ A_{rec} $, where the area of each photodiode equals to $ A_m = \frac{A_{rec}}{M} $ with radius $ r_{m} $ , $ m \in M $.  Note that,  each photodiode points to a distinct direction to guarantee wide FoV, and its orientation vector  is determined by the elevation $\theta^{[k,m]}$ and azimuth $\alpha^{[k,m]}$ angles on the x-y plane, as follows 
\begin{equation}
\begin{split}
\hat{\mathbf{n}}^{[k,m]}=&
\left[ \sin\left(\theta^{[k,m]} \right)\cos\left(\alpha^{[k,m]}\right), \,\, \right .\\
 &\phantom{\{} \left . \sin\left(\theta^{[k,m]} \right)\sin\left(\alpha^{[k,m]}\right), \,\, \cos\left(\theta^{[k,m]} \right) \right],
\end{split}
\end{equation}
At this point, the irradiance angle $ \phi_{t}^{[k]} $ of the VCSEL towards user $ k $ can be  determined by
$\phi_{t}^{[k]}=\cos^{-1} \left(\frac{\hat{\mathbf{n}}_{t} \cdot {\mathbf{d_{t,k}}}}{\Vert{\mathbf{d_{t,k}}}\Vert} \right)$, and the incidence angle $ \psi_{t}^{k} (m) $ can be expressed as  $\psi_{t}^{k} (m)=\cos^{-1} \left(\frac{\hat{\mathbf{n}}^{[k,m]}  \cdot {\mathbf{-d_{t,k}}}}{{\Vert{\mathbf{d_{t,k}}}\Vert}} \right)$, where $ {\mathbf{d_{t,k}}}= \mathbf{P}_{t}- \mathbf{P}_k $ is  the distance vector determined according to the locations of the transmitter and user, $ \mathbf{P}_{t} $ and  $ \mathbf{P}_k $, respectively. In this context, the spatial distribution of the intensity of the transmitter is a function of $ {\Vert{\mathbf{d_{t,k}}}\Vert} $ and $ \phi $, i.e., $ I_{t}({\Vert{\mathbf{d_{t,k}}}\Vert}, \phi) $. Thus, the optical channel between photodiode $ m $ of user $ k $ and  the VCSEL is given by 
\begin{multline}
h^{[t,m]}= I_{t}({\Vert{\mathbf{d_{t,k}}}\Vert}, \phi) {A_m} G_{m} \cos (\psi_{t}^{k} (m)) $rect$ \left(\frac{\psi_{t}^{k} (m)}{\Psi_{F}}\right)\\
= \frac{2  \cos (\psi_{t}^{k} (m)) P_{tr} {A_m} G_{m}}{\pi W^{2}({\Vert{\mathbf{d_{t,k}}}\Vert} \cos \phi_{t}^{[k]})}\\ \mathrm{exp}\left(-\frac{2 {\Vert{\mathbf{d_{t,k}}}\Vert}^{2} \sin^{2} \phi_{t}^{[k]}}{W^{2}({\Vert{\mathbf{d_{t,k}}}\Vert} \cos \phi_{t}^{[k]})}\right) $rect$ \bigg(\frac{\psi_{t}^{k} (m)}{\Psi}\bigg),
\end{multline}
where $ G_{m} $ is the  gain of the photodiode, and $ \Psi_{F} $ is  the  field of view  where $ $rect$ \bigg(\frac{\psi_{t}^{k} (m)}{\Psi}\bigg)=1 $  if $ 0 \leq \psi_{t}^{k}  \leq\Psi_{F} $, or it equals zero otherwise. As considered in the majority of the previous works on OWC, we neglect  the diffuse components of the optical channel considering the fact that the biggest portion of the optical power received is due to the line of sight (LoS) components\cite{mo6963803,AA19901111,owc-sp22}.
\subsection{Eye safety}
The use of the VCSEL or any IR laser as  a transmitter in indoor environments might be harmful to the human eye where a power constraint must be imposed to ensure that the transmit power of the VCSEL is safe at any position in  the room. According to \cite{mo6963803}, the maximum
permissible exposure  to the power of the laser is set as a function of  the wavelength and the exposure time.  Therefore,  the exposure level at the most hazard potion in the room must be less than or equal to  the maximum
permissible exposure, more details are in the appendix. At this point, we limit the transmit power of the VCSEL to any value within the range $ [P_{\min}, P_{\max}] $, where $ P_{\min} $ is the minimum power to turn the VCSEL on, and $ P_{\max} $ is the maximum permissible power. { It is worth mentioning that the power range corresponds to the maximum and minimum input currents for the VCSEL, $ [I_{\min}, I_{\max}]  $, where  the
amplitude of the transmitted signal must be less than or equal to $\min (I_{Dc}-I_{\min},I_{\max}-I_{Dc})$.}

In this work, each optical AP $ l \in \mathcal{L} $ contains $ L_{v} \times L_{v} $ VCSELs, and therefore, $ P_{\min} \leq \frac{P_l}{L_{v} \times L_{v}} \leq P_{\max}$, where $ P_{l} $ is the total transmit power of the AP, ensuring that all the VCSELs transmit the same power, which complies with the eye safety regulation as determined in the appendix.

\begin{figure}[t]
\begin{center}\hspace*{0cm}
\includegraphics[width=1\linewidth]{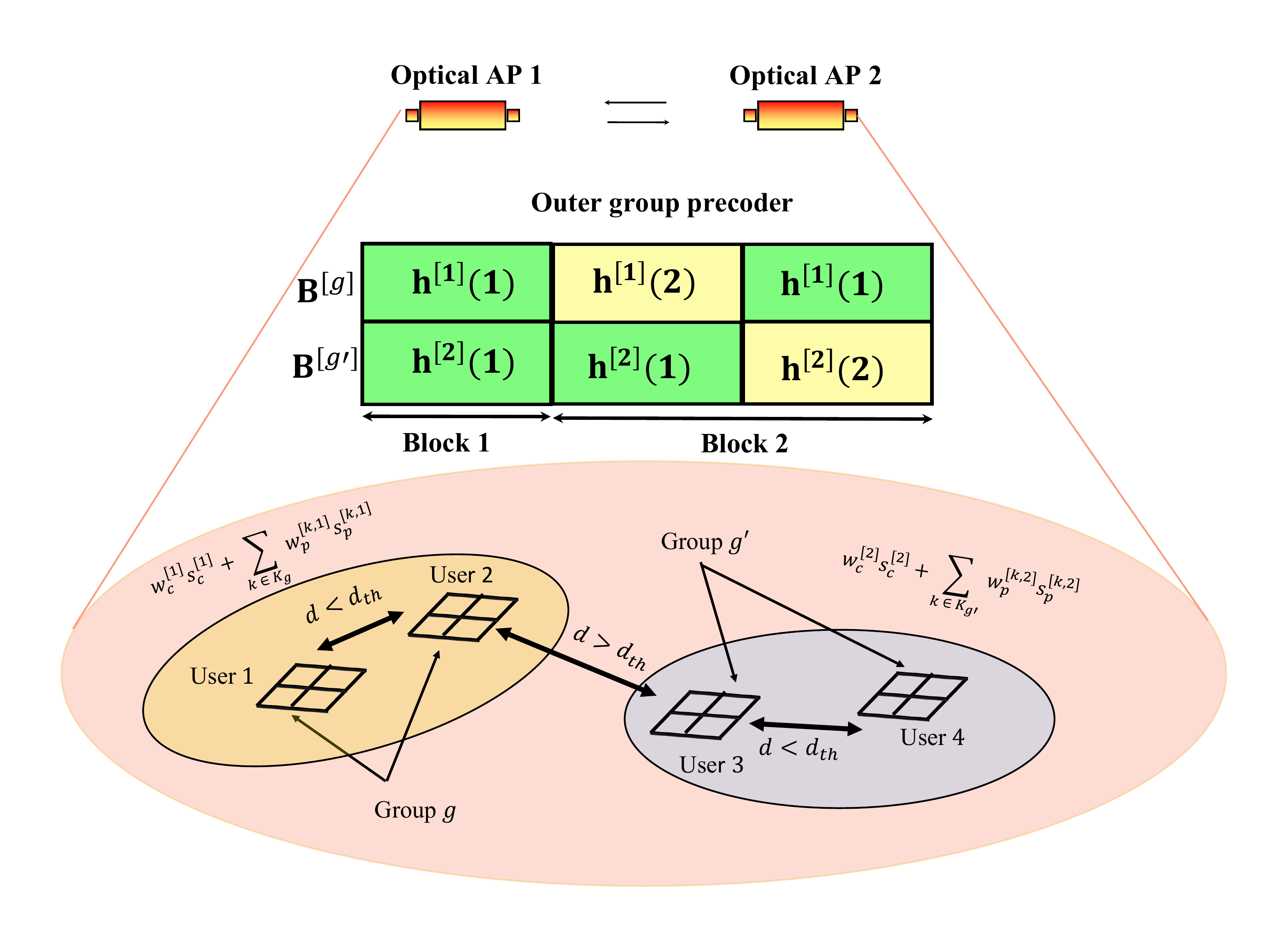}
\end{center}
\vspace{-2mm}
\caption{An OWC use case where $ L=2 $ APs serving $ G=2 $ groups, each group $ g $ with $ K_{g}=2 $ users. The users of each group receive the same signal.}\label{Fig4}
\vspace{-5mm}
\end{figure}
\section{ A Novel Rate Splitting-based  Interference Management Scheme}
\label{sec:CSIRS}
The conventional  RS scheme derived in \cite{7434643, HRS} is suitable for RF networks due to the large coverage area of the RF AP. However, RS is subject to performance-degradation in high density OWC networks in which a high number of optical APs are deployed to expand the coverage and serve multiple active users simultaneously. In other words, users in an OWC network experience severe ICI if  the transmission among the optical APs is not coordinated properly. Therefore, we design a novel and general framework transmission scheme, namely BIA-RS transmission, which smartly combines the features of both BIA and RS resulting in a robust interference management scheme. Basically, the BIA methodology is followed to  design an outer precoder  that coordinates the transmission among multiple APs, turning ICI into useful information, and allocates non-orthogonal resources to users divided into multiple groups in a way that guarantees enough dimensions for the users to measure inter-group interference. Therefore, parallel RS schemes  can work simultaneously in multiple groups to manage intra-group interference and maximize the performance of the network. It is worth mentioning that traditional BIA schemes have low performance in OWC networks as the size of the transmission block increases due to the requirements of the channel coherence time, which is determined according to the number of APs and users \cite{8636954,9064520,9500371} \cite{8935164}. However, the transmission block of BIA-RS is given by the number of APs and groups, and therefore, its application relaxes such requirements. In this section, we first report  the design of the BIA-RS scheme for a use-case, and then, the general case is considered to derive the sum rate of the network.

\subsection{Motivational example of BIA-RS}
For illustrative purposes, we first derive the mathematical expressions of the BIA-RS scheme for a toy OWC example that is composed of $ L=2 $ optical AP serving $ K=4 $ users forming $ G=2 $ groups, each group with $ K_{g} $ users, as  shown in Fig.~\ref{Fig4}. In this context, the methodology of BIA explained in the previous section is followed to coordinate the transmission and  determine the precoding matrix of each group denoted by $ \mathbf{B^{[g]}} $, i.e., the users of each group use the same outer precoding matrix, such that the interference among the groups is aligned. Therefore, the transmitted signal considering the application of RS within each group can be expressed as 
{
\begin{multline}
\label{bia-rs}
\mathbf{X}= \mathbf{B}^{[1]} \bigg(\rho\sqrt{P_{c}^{[1]}} \mathbf{w}_{c}^{[1]} \mathbf{s}_{c}^{[1]}+ \rho\sum^{K_g}_{k=1}\sqrt{P_{p}^{[k,1]}}\mathbf{w}_{p}^{[k,1]} \mathbf{s}_{p}^{[k,1]}\bigg)
\\ +\mathbf{B}^{[2]} \bigg(\rho\sqrt{P_{c}^{[2]}} \mathbf{w}_{c}^{[2]} \mathbf{s}_{c}^{[2]} + \rho\sum^{K_g}_{k=1}\sqrt{P_{p}^{[k,2]}}\mathbf{w}_{p}^{[k,2]} \mathbf{s}_{p}^{[k,2]}\bigg) \\+\rho I_{Dc} \times \mathbf{1}_{2\times1}, 
\end{multline}}where $ {P_{c}^{[g]}} $ is the power of the common message intended to group $ g $, $ \mathbf{w}_{c}^{[g]} $   is the unit-norm precoding vector of the common
message intended to group $ g $ and $ \mathbf{s}_{c}^{[g]} $ denotes the common message intended to  users 1 and 2 belonging to group $ g $. Moreover, $ P_{p}^{[k,g]} $, $ k \in \mathcal{K}_{g} $, is the power allocated to the private message intended to user $ k $  belonging to group $ g $, $ \mathbf{w}_{p}^{[k,g]} $ is the unit-norm precoding vector of the private message intended to user $ k \in \mathcal{K}_{g}$, and
 $ \mathbf{s}^{[g]}_{p} =[\mathbf{s}_{p}^{[1,g]}, \mathbf{s}_{p}^{[2,g]}] \in \mathbb{C}^{2} $ is the data vector of the private messages  intended
to the user 1 and 2 of group $ g $. In this example, the transmission from two transmitters to two groups must occur over a predefined  transmission block comprising  three time slots in accordance to the BIA methodology as in \eqref{bia-rs}, where  the users of each group switch their channel pattern over  two time slots to receive information, while devoting a certain time slot to measure  the interference received due to the transmission to the adjacent group. Therefore, the BIA-based outer precoding matrices for groups 1 and 2  denoted by  $ \mathbf{B}^{[1]} $ and $ \mathbf{B}^{[2]} $, respectively,  can be expressed as      
\begin{equation}
\label{matrix}
\mathbf{B}^{[1]}= \begin{bmatrix}
\mathbf{I_2}\\
\mathbf{I_2}\\
\mathbf{0_2}\\
\end{bmatrix}, ~  \mathbf{B}^{[2]}= \begin{bmatrix}
\mathbf{I_2}\\
\mathbf{0_2}\\
\mathbf{I_2}\\
\end{bmatrix},
\end{equation}
It is worth pointing out that these outer precoding matrices are given by  0 and
1 following the structure of the BIA transmission block as shown in Fig.~\ref{Fig4}.
{Focussing on user 1 belonging to group 1, without loss of generality, and omitting the received DC part to shorten the equation, the received signal can be expressed as   
\begin{multline}
\label{recived signal}
\mathbf {y}^{[1,1]}= \\\zeta\underbrace{\begin{bmatrix}
\mathbf{h}^{[1,1]}(1)^{T}\\ 
\mathbf{h}^{[1,1]}(2)^{T}\\
\mathbf{0}_{2,1}^{T}\\
\end{bmatrix}}_{\text{rank=2}}
\bigg(\rho\sqrt{P_{c}^{[1]}} \mathbf{w}_{c}^{[1]} \mathbf{s}_{c}^{[1]}+ \rho\sum^{K_g}_{k=1}\sqrt{P_{p}^{[k,1]}}\mathbf{w}_{p}^{[k,1]} \mathbf{s}_{p}^{[k,1]}\bigg)
\\+ \zeta \underbrace{\begin{bmatrix}
\mathbf{h}^{[1,1]}(1)^{T}\\
\mathbf{0}_{2,1}^{T}\\
\mathbf{h}^{[1,1]}(1)^{T} \\
\end{bmatrix}}_{\text{rank=1}} \bigg(\rho\sqrt{P_{c}^{[2]}} \mathbf{w}_{c}^{[2]} \mathbf{s}_{c}^{[2]}+\rho \sum^{K_g}_{k=1}\sqrt{P_{p}^{[k,2]}}\mathbf{w}_{p}^{[k,2]} \mathbf{s}_{p}^{[k,2]}\bigg)
\\+\begin{bmatrix}
{z}^{[1,1]}(1)\\ 
{z}^{[1,1]}(2)\\
{z}^{[1,1]}(3)\\
\end{bmatrix},
\end{multline}}where $ \mathbf {y}^{[1,1]}= \text{col}\{{y}^{[1,1]}(\kappa)\}^{3}_{\kappa=1} $. It can be seen from equation \eqref{recived signal} that the desired information of user 1 appears into a full rank matrix $ [1~ 1~ 0]^{T} $, while the interference caused  by the transmission to the users belonging to group 2 is received  into one dimension $ [1~ 0~ 1]^{T} $. Thus, the desired information is decodable after measuring and canceling inter-group interference. Note that, user 1 receives  the desired information over the first time slot polluted with interference  and the second time slot, which is an interference-free time slot. In other words, the first and second time slots form an alignment block over which the users of group 1 receive information. As a consequence, user 1 devotes the third time slot to measure and cancel the interference received over the first time slot  at the cost of noise enhancement. Thus, the received signal of user 1 belonging to group 1  after  inter-group interference cancellation can be written as  
{
\begin{multline}
\label{recived signal 2}
\tilde{\mathbf{y}}^{[1,1]}= \rho\zeta{\begin{bmatrix}
\mathbf{h}^{[1,1]}(1)^{T}\\ 
\mathbf{h}^{[1,1]}(2)^{T}\\
\end{bmatrix}}
\bigg(\sqrt{P_{c}^{[1]}} \mathbf{w}_{c}^{[1]} \mathbf{s}_{c}^{[1]}+\sqrt{P_{p}^{[1,1]}}\mathbf{w}_{p}^{[1,1]} \\\mathbf{s}_{p}^{[1,1]}+\sqrt{P_{p}^{[2,1]}}\mathbf{w}_{p}^{[2,1]} \mathbf{s}_{p}^{[2,1]}\bigg)
+\begin{bmatrix}
{z}^{[1,1]}(1)-{z}^{[1,1]}(3)\\ 
{z}^{[1,1]}(2)\\
\end{bmatrix}.
\end{multline}}
It can be easily seen that the received signal of user 1 is free of inter-group interference with an increase in noise. Interestingly, $\mathbf{h}(1) $ and  $\mathbf{h}(2) $ are linearly independent channel responses, and therefore, the information of user 1 can be decoded after canceling intra-group interference by  applying the RS technique. In  \eqref{recived signal 2},
the message of user 1 is split into common and private messages, $ \mathbf{s}_{c}^{[1]} $ and $ \mathbf{s}_{p}^{[1,1]}  $. In this case, user 1 follows the same procedure of conventional RS to decode its private message by decoding first the common message $ \mathbf{s}_{c}^{[1]} $, while treating  the private messages $ \mathbf{s}_{p}^{[1,1]}  $ and $ \mathbf{s}_{p}^{[2,1]}  $ as noise. After that, the decoded common message is  eliminated by applying SIC, giving user 1 the opportunity to decode its interference-free private message. Note that, user 2 belonging to group 1 follows the same procedure to decode the private message $ \mathbf{s}_{p}^{[2,1]}  $, as well as the other users belonging to group 2.
\subsection{General framework of BIA-RS}
We define the BIA-RS transmission scheme as a general framework that is applicable for different scenarios. Thus, we consider an OWC system composed of $ L $ optical APs serving $ K $ users arranged  into
 a set of groups $ G $, and each group contains $ K_{g} $ users. The proposed BIA-RS scheme has the ability to coordinate the transmission among $ L $ optical APs and  deliver interference-free signals for the users in the network  according the the following
criterion; {\it the methodology of BIA is used to create a transmission block comprising a number of alignment blocks  allocated to each group  by the whole set of the transmitters. In contrast to the conventional BIA schemes, the number of alignment blocks allocated to each group is given by  $\ell=\{1, \dots, (L-1)^{G-1}\} $, and therefore, a total of $ G\times (L-1)^{G-1} $  alignment blocks are built for all the groups over the transmission block. To satisfy  the  rules of constructing these alignment blocks while guarantee the ability to measure inter-group interference by the users belonging to each group, the transmission block must comprise $ (L-1)^{G}+ G \times(L-1)^{G-1} $ time slots.} 
Once the construction of the transmission block is determined, the BIA-based outer precoding matrix of each group $ \mathbf{B}^{[g]} $ can be designed by arranging each alignment block $ \ell $ into a column in the precoding matrix, and the rows corresponding to the  time slots over which the transmission occurred are given by $ L \times L $ identity matrices. For instance, in equation \eqref{matrix}, the precoding matrix $  \mathbf{B}^{[1]}  $ contains 2 $ \times $ 2 identity matrices in the first and second rows corresponding to the first and second time slots that form the only alignment block allocated to group 1, which is determined  according to the number of transmitters and groups in that use case. It is worth mentioning that the outer precoding matrices for all the groups can be determined in this way following the structure of the BIA transmission block predefined at the transmitters. Considering that the users belonging to the group use the same outer precoding matrix, the transmitted signal for the general case can be expressed as
{
\begin{multline}
\mathbf{X}= \sum^{G}_{g=1}\mathbf{B}^{[g]} \bigg(\rho\sqrt{P_{c}^{[g]}} \mathbf{w}_{c}^{[g]} \mathbf{s}_{c}^{[g]}+\rho\sqrt{P_{p}^{[k,g]}}\mathbf{W}_{p}^{[g]} \mathbf{s}_{p}^{[g]}\bigg)\\+\rho I_{Dc} \times \mathbf{1}_{L\times1}, 
\end{multline}}
Note that, $ \mathbf {X}= \text{col} \{\mathbf{x}(\kappa)\}_{\kappa=1}^{\mathcal{V}} $, where $ \mathcal{V} $ is the total number of time slots that form the transmission block of BIA,   $ \mathbf{W}_{p}^{[g]}= \bigg[ \mathbf{w}_{p}^{[1,g]}, \dots, \mathbf{w}_{p}^{[k,g]}, \dots, \mathbf{w}_{p}^{[K_{g},g]} \bigg] $ is the inner precoding vector that contains  the inner precoding matrices of the private messages intended to the $ K_{g} $ users belonging to group $ g $, and $ \mathbf{s}_{c}^{[g]} =\text{col} \bigg\{\mathbf{s}_{c, \ell}^{[g]} \bigg\}_{\ell}^{(L-1)^{G-1}} $ contains the common messages transmitted to group $ g $, $ g\in \mathcal{G}$, during the entire transmission block, where  $ \mathbf{s}_{c, \ell}^{[g]} $ is the common message transmitted over a certain alignment block $ \ell  $. Moreover, $ \mathbf{s}_{p}^{[g]}= \bigg[ \mathbf{s}_{p}^{[1,g]}, \dots, \mathbf{s}_{p}^{[k,g]}, \dots, \mathbf{s}_{p}^{[K_{g},g]} \bigg] $, where $ \mathbf{s}_{p}^{[k,g]} =\text{col} \bigg\{\mathbf{s}_{p,\ell}^{[k,g]}\bigg\}_{\ell}^{(L-1)^{G-1}} $ contains the private messages transmitted to user $ k $, $ k\in \mathcal{K}_{g}$, over all the alignment blocks allocated to group $ g $, where $ \mathbf{s}_{p,\ell}^{[k,g]} $   is the private message transmitted over a certain alignment block $ \ell $. For detailed  mathematical explanation on the allocation of alignment blocks  to multiple groups over the BIA transmission block, we refer readers to \cite{GWJ11}. Note that, power allocation in BIA-RS transmission is addressed in the next section. 

The BIA-based outer precoding matrix  of each group gives the opportunity for $ K_{g} $ users  to measure and cancel  the interference received due to serving other $ K_{g'} $ users belonging to adjacent group $ g' $, $ g' \neq g $. At this point, the $ K_g $ users of each group $ g $ are subject to intra-group interference.  Without loss of
generality,  the received signal of user $ k\in \mathcal{K}_{g} $ polluted by intra-group interference can be written as 
{
\begin{multline}
\tilde{\mathbf{y}}^{[k,g]}=  \rho\zeta{\mathbf{H}}^{[k,g]}\bigg( \sqrt{P_{c}^{[g]}}  \mathbf{w}_{c}^{[g]} \mathbf{s}_{c}^{[g]}+ \sqrt{P_{p}^{[k,g]}} \mathbf{w}_{p}^{[k,g]} \mathbf{s}_{p}^{[k,g]}\\ \underbrace{+ \sum^{K_g}_{k'\neq k}\sqrt{P_{p}^{[k',g]}}  \mathbf{w}_{p}^{[k',g]} \mathbf{s}_{p}^{[k',g]}}_{\text{intra-group interference}}\bigg)+ \tilde{\mathbf{z}}^{[k,g]}.
\end{multline}}
Following the conventional RS procedure in canceling intra-group interference, user $ k $ can decode its desired information at the cost of noise enhancement. However, the proposed BIA-RS scheme results in low noise compared with traditional RS due to the fact that each user measures and cancels the private messages sent to $ K_{g} < K $ users regardless of all other users in the network.

We derive now the achievable sum rate of the BIA-RS transmission scheme for the general case taking into account the distinct
features of the optical signal and the optical power constraints. Focussing on the use of RS within each group, the message of user $ k $ is divided into common and private messages. These messages are decodable, and user $ k $ belonging to group $ g $ decodes first the common message treating all the private messages as noise, and therefore, { the SINR of the common message is upper bounded according to \cite{DC7378985}, and can be expressed as  
\begin{equation}
\label{snrc}
\gamma_{c}^{[k,g]}=\frac{c\rho^{2} {\zeta}^{2} P^{[g]}_{c}\left|\mathbf{w}^{[g]}_{c}\right|^{2}}{c\rho^{2} {\zeta}^{2}\sum_{k=1}^{K_g} P^{[k,g]}_{p}\left|\mathbf{w}^{[k,g]}_{p}\right|^{2}+{\sigma_{z}}^{2}}.
\end{equation}
where $ c=\min\{\frac{1}{2\pi e}, \frac{e {I^{2}_{DC}}}{2 {I^{2}_{\max}}\pi}\}$ where $e$ is the Euler number. Note that, the deriving input current to the VCSEL is   ${I_{DC}}=\frac{I_{\max}}{2}$. Hence, $c=\frac{1}{2\pi e} $. }Subsequently, user $ k $ decodes its private message treating the private messages intended to other users belonging to group $ g $  as noise. That is, the SINR of the private message upper bounded according to \cite{DC7378985} is given by  
\begin{equation}
\label{snrp}
\gamma^{[k,g]}_{p}=\frac{c\rho^{2} {\zeta}^{2} P^{[k,g]}_{p}\left|\mathbf{w}^{[k,g]}_{p}\right|^{2}}{c\rho^{2} {\zeta}^{2}\sum^{K_g}_{k' \neq k} P^{[k',g]}_{p}\left| \mathbf{w}^{[k',g]}_{p}\right|^{2}+{\sigma_{z}}^{2}}.
\end{equation} Considering the full procedure of the BIA-RS scheme in managing interference among multiple groups in the network, {the sum rate of the common messages is given by 
\begin{equation}
\label{ratec}
{R_{c}}=\sum_{g=1}^{G} b_{ab} \times  \mathbb{E}\left[ \log \det\left( \mathbf{I} +  \gamma_{c}^{[k,g]} \mathbf{H^{[k,g]}}\mathbf{H^{[k,g]^{H}}}  \mathbf{R_{\tilde{z}}}^{-1}\right) \right],
\end{equation} }
where $  b_{ab}$ is the ratio of the alignment blocks allocated  uniformly among the groups. For $ L $ transmitters serving $ G $ groups, it is given by $ \frac{1}{G+L-1} $. Moreover, $ \mathbf{R_{\tilde{z}}} $ is  the covariance matrix that results from subtracting the information intended to $ G-1 $ groups.
On the other hand, {the sum rate of the private messages intended to $ K_g $ users belonging to each group can be expressed as 
\begin{equation}
\label{ratep}
R_{p}^{[g]}= \sum^{K_{g}}_{k=1} b_{ab} \times  \mathbb{E}\left[ \log \det\left( \mathbf{I}  +  \gamma_{p}^{[k,g]} \mathbf{H^{[k,g]}}\mathbf{H^{[k,g]^{H}}}  \mathbf{R_{\tilde{z}}}^{-1}\right)\right].
\end{equation} }
As a result, the overall sum rate of the proposed scheme is equal to  $ R_{\tiny{\text{BIA-RS}}}= R_{c}+ \sum_{g=1}^{G} R_{p}^{[g]} $.
\section{Power Allocation in BIA-RS transmission}
\label{sec:power}
The proposed BIA-RS scheme eliminates ICI and  inter-group interference following the construction of the BIA transmission block. However, the traditional BIA transmission block considers uniform power allocation among the groups, which causes high power consumption if the  users of a certain group do not use their power fully. Besides, the powers of the common and private messages must be determined in an optimal fashion to ensure the principles of RS. In this context, we derive a power allocation optimization problem  for data rate-maximization to efficiently use the power budget among the messages of BIA-RS intended to the  users belonging to different groups.  
  
\subsection{Problem formulation}
Considering the number of groups $ G $ and the common and private messages sent to each $ K_g $ users, the sum  rate of the proposed BIA-RS can be expressed as 
\begin{equation}
U (R_{\tiny{\text{BIA-RS}}})=  \sum^{G}_{g} \bigg( {R^{[g]}_{c}} (P^{[g]}_{c}, P^{[k,g]}_{p})+ \sum^{K_{g}}_{k=1} {R^{[k,g]}_{p}} ( P^{[k,g]}_{p} ) \bigg),
\end{equation}
Note that, $  {R^{[g]}_{c}} (P^{[g]}_{c},P^{[k,g]}_{p}) $ and $ {R^{[k,g]}_{p}} ( P^{[k,g]}_{p} ) $ can be easily derived from equations \eqref{ratec} and \eqref{ratep}, respectively.  Note that, each group  is composed of a unique set of users, and therefore, the overall data rate of the network can be maximized by ensuring higher sum rate within each group. In this context, we formulate an  optimization problem  that maximizes the minimum sum rate of the $ K_g $ users  belonging to each group through controlling power allocation  among the common and  private messages as follows:  

\begin{equation}
\label{OP1}
\begin{aligned}
\max_{p} \quad & \bigg\{ \min_{g \in \mathcal{G}}   {R^{[g]}_{\text{sum}}} (P^{[g]}_{c},P^{[k,g]}_{p})  \bigg\}  \\ \\
\textrm{s.t.} \quad & {R^{[g]}_{\text{sum}}} (P^{[g]}_{c},P^{[k,g]}_{p}) \geq R^{[g]}_{min},~~~~\\
 \quad & P^{[g]}_{c}+\sum_{k \in \mathcal{K}_{g}} P^{[k,g]}_{p}  \leq P^{[g]}_{max}, ~~~~~~~~~~\forall g \in \mathcal{G},\\
~~~~~~~~~~~\quad &  P^{[k,g]}_{p} \leq P^{T}_{p},~~~~~~~~~~~~~~~~~~~~~~~~~~ \forall k \in \mathcal{K}_{g},\\ 
\quad &  P^{[k,g]}_{p} \geq 0, \sum_{g\in \mathcal{G}} P^{[g]}_{max}\leq P_{T},~~~~ k \in \mathcal{K}_{g}, g \in \mathcal{G},
\end{aligned}
\end{equation}
where $ {R^{[g]}_{\text{sum}}} (P^{[g]}_{c},P^{[k,g]}_{p}) =  \bigg( {R^{[g]}_{c}} (P^{[g]}_{c}, P^{[k,g]}_{p})+ \sum^{K_{g}}_{k=1} {R^{[k,g]}_{p}} ( P^{[k,g]}_{p} )\bigg)$. The optimization problem in \eqref{OP1} is defined as max-min fractional program under several constraints with a particular structure. It is  classified as concave-convex fractional
program that has high complexity \cite{ReP}. The first constraint guarantees that the sum rate of the users belonging to a given group $ g $ is higher than the minimum data rate $ R^{[g]}_{min} $ required to ensure high quality of service. The second constraint controls the total power allocated to the common and private messages intended to $ K_g $ users, which must be less than  or equal to the maximum power dedicated to each group $ P^{[g]}_{max} $ to easy measure  the interference caused by the transmission to the other groups in the network \cite{7437435}. The third constraint limits the power allocated to the private message intended to a certain user, $ k\in \mathcal{K}_{g} $, where $ P^{T}_{p}$ is the maximum power devoted to send each of the  $ K_g $ private messages in order to guarantee enough power for the common message needed to manage  intra-group interference \cite{Kh9500371}. The last constraint defines the feasible region of the optimization problem, as well as ensures that the total power consumed by $ G $ groups is less than or equal to the maximum power consumption $  P_{T}=\sum^{L}_{l=1} P_{l} $ allowed in the network, which is calculated according to our system model that guarantees eye safety.
\subsection{Sub-optimal solution}
The optimization problem in \eqref{OP1} can be solved using a parametric
approach \cite{ReP}, where it can be transformed  into  a convex
optimization problem at a given parameter value. Thus, the new  objective function can be written as
\begin{equation}
\label{CSIopt2}
f (\Upsilon)= \max_{p}  \bigg\{ \min_{g \in \mathcal{G}}  \Big\{ {R^{[g]}_{\text{sum}}} (P^{[g]}_{c},P^{[k,g]}_{p})- \Upsilon  \Big\} \bigg\},
\end{equation}
where $ \Upsilon= P^{[g]}_{T} \delta $, considering $ P^{[g]}_{T}= P^{[g]}_{c}  +\sum_{k \in \mathcal{K}_{g}} P^{[k,g]}_{p}  $ as  the total power  consumed by group $ g $. Moreover,  $ \delta= \Big(\min\limits_{g \in \mathcal{G}}  \Big\{ {R^{[g]}_{\text{sum}}} (P^{[g]}_{c},P^{[k,g]}_{p})/ P^{[g]}_{T} \Big\} \Big)$ is a non-negative parameter where at its optimal value, the power is allocated among the common and  private messages intended to $ K_g $ users maximizing the minimum sum rate of group $ g $. According to \cite{opop}, equation \eqref{CSIopt2} can be optimally determined under the constraints in equation \eqref{OP1} through finding a
root of $f (\Upsilon)=0$ using Dinkelbach-type algorithm. From equations \eqref{OP1} and \eqref{CSIopt2}, the optimization problem can be reformulated as  


\begin{equation}
\label{OP10}
\begin{aligned}
\max_{p} \quad & \bigg\{ \min_{g \in  \mathcal{G}}  \Big\{ {R^{[g]}_{\text{sum}}} (P^{[g]}_{c},P^{[k,g]}_{p})- \Upsilon  \Big\} \bigg\} \\
\textrm{s.t.} \quad & \min_{g \in \mathcal{G}}  \Big\{ {R^{[g]}_{\text{sum}}} (P^{[g]}_{c},P^{[k,g]}_{p})- \Upsilon  \Big\} \\ & ~~~~~~~~~~~~~~~~~~~~~~\leq {R^{[g]}_{\text{sum}}} (P^{[g]}_{c},P^{[k,g]}_{p})- \Upsilon,\\
\quad & \text{ (1-4) Constraints in \eqref{OP1}}.
\end{aligned}
\end{equation}
Interestingly, the optimal solution of the optimization problem in \eqref{OP10} can be equivalently found  using a distributed algorithm via Lagrangian decomposition, i.e., using Lagrangian multipliers \cite{7054502,9521837}. Given that, The
Lagrangian function of  \eqref{OP10} considering other groups in the network can be written as

\begin{equation}
\label{lag}
\mathcal{F}= \min_{g \in \mathcal{G}}  \Big\{ {R^{[g]}_{\text{sum}}} (P^{[g]}_{c},P^{[k,g]}_{p})- \Upsilon  \Big\} \left(  1-\sum_{g \in \mathcal{G}} \lambda_{g} \right) + \sum_{g \in \mathcal{G}} \mathcal{F}_{g},
\end{equation}
where 
\begin{multline}
\label{fg}
\mathcal{F}_{g}=  \Big\{(\lambda_{g}+ \xi_{g}) {R^{[g]}_{\text{sum}}} (P^{[g]}_{c},P^{[k,g]}_{p})\\- (\delta \lambda_{g}+ \nu_{g}) (P^{[g]}_{c}+\sum_{k \in \mathcal{K}_{g}} P^{[k,g]}_{p}) \\+ \beta^{[k]}_{g} (P^{T}_{p} - P^{[k,g]}_{p}) - \xi_{g}  R^{[g]}_{min} + \nu_{g} P^{[g]}_{max}  \Big\}, 
\end{multline}
and $ \lambda_{g} $, $ \xi_{g} $, $ \nu_{g} $ and $ \beta^{[k]}_{g} $ are Lagrangian multipliers corresponding to  the constraints in equation  \eqref{OP10}, respectively.  Therefore,
the optimal values of the power allocated to the private messages in the network can be equivalently found through maximizing $ \mathcal{F} $ in equation  \eqref{lag}. 
However,  power allocation-based data rate-maximization in RS transmission  is a complex task due to  the fact that  the powers of the common and private messages are coupled (see equations \eqref{snrc} and \eqref{snrp}). In this context, we consider  sending  the private messages to the $ K_g $ users at the highest permissible power $ P^{T}_{p} $, and therefore, the interference term in the common data rate can be managed as constant, decoupling $ P^{[g]}_{c} $  and $ P^{[k,g]} $. Therefore, according to  the Karush-Kuhn-Tucker (KKT) conditions \cite{100020202}, the partial derivatives $ \frac{\partial \mathcal{F}_{g} }{\partial P^{[k,g]}_{p} } $ and  $ \frac{\partial \mathcal{F}_{g} }{\partial P^{[g]}_{c} } $ 
 are defined as monotonically decreasing functions with respect to the power  $  P^{[k,g]}_{p}  $ allocated  to user $ k $ belonging to group $ g $ and to the power $ P^{[g]}_{c} $ allocated to the common message received by all the $ K_g $ users, respectively. That is, if the partial derivatives $ \frac{\partial \mathcal{F}_{g} }{\partial P^{[k,g]}_{p} } \vert _{{P^{[k,g]}_{p}} =0}\leq 0 $ and $ \frac{\partial \mathcal{F}_{g} }{\partial P^{[g]}_{c} } \vert _{P^{[g]}_{c} =0}\leq 0 $, the optimum powers $ P^{*[k,g]}_{p}$ and $ P^{*[g]}_{c}$  equal zero. Moreover, if the partial derivatives $ \frac{\partial \mathcal{F}_{g} }{\partial P^{[k,g]}_{p} } \vert _{{P^{[k,g]}_{p}}=1}\geq 0 $ and $ \frac{\partial \mathcal{F}_{g} }{\partial P^{[g]}_{c} } \vert _{{P^{[g]}_{c}}=1}\geq 0 $, the optimum powers $ P^{*[k,g]}_{p} $ and $ P^{*[g]}_{c}$ equal one.  Otherwise, the optimum powers  allocated to the common and private messages can be found as follows

\subsubsection{Optimality at  $ \beta^{[k]}_{g} $ and $ \nu_{g} $ values }
At  given values for $ \Upsilon $, $ \lambda_{g} $ and $ \xi_{g},  $ $ \forall g \in \mathcal{G} $, $ P^{*[k,g]}_{p} $ can be calculated for each user, $ k \in \mathcal{K}_{g} $, belonging to group $ g $ assuming the other users, $ k' \neq k,  k' \in K_{g},  $ served at $ P^{T}_{p} $ as follows

\begin{multline}
\label{grad2}
\frac{\partial \mathcal{F}_{g} }{\partial P^{[k,g]}_{p} }=  (\lambda_{g}+ \xi_{g}) \frac{\partial {R^{[k,g]}_{p}} (P^{[k,g]}_{p}) }{\partial P^{[k,g]}_{p} }
- (\delta \lambda_{g}+ \nu_{g}+\beta^{[k]}_{g})=0.
\end{multline}
While $ P^{*[g]}_{c} $ can be calculated for each group, $ g \in \mathcal{G} $,  as follows
\begin{multline}
\label{gradcom}
\frac{\partial \mathcal{F}_{g} }{\partial P^{[g]}_{c} }=  (\lambda_{g}+ \xi_{g}) \frac{\partial {R^{[k,g]}_{c}} ( P^{[g]}_{c}, P^{[k,g]}_{p}) }{\partial  P^{[g]}_{c} }
- (\delta \lambda_{g}+ \nu_{g})=0.
\end{multline}
Therefore, equation \eqref{fg} can be modified taking into consideration the optimum values of the private and common messages from \eqref{grad2} and \eqref{gradcom}. 
At this point, the gradient projection method can be applied to solve the dual problem, then, updating  the Lagrangian multipliers   $ \beta^{[k]}_{g} $ and $ \nu_{g} $  as follows 

\begin{equation}
\label{var1}
  \beta^{[k]}_{g} (\kappa)= \left[\beta^{[k]}_{g}(\kappa-1)-\epsilon_{1}(\kappa-1) \left(P^{T}_{p}-P^{[k,g]}_{p} (\kappa-1) \right) \right]^{+},
\end{equation} 
\begin{multline}
\label{var2}
\nu_{g} (\kappa)= \bigg[\nu_{g}(\kappa-1)-\epsilon_{2}(\kappa-1) \Big(P^{[g]}_{max} \\- \Big(P^{[g]}_{c} +\sum_{k \in \mathcal{K}_{g}} P^{[k,g]}_{p} (\kappa-1)\Big)\Big) \bigg]^{+},
\end{multline}
where $ \kappa$ denotes the iteration of the gradient algorithm, and  $ [.]^{+} $ is a projection on the positive orthant
to take into account the fact that we have $ \beta^{[k]}_{g} , \nu_{g} \geq 0 $. Moreover, $ \epsilon_{1} $ and $ \epsilon_{2} $  are sufficient small step sizes at a given iteration $ (\kappa-1)$  that are taken in the direction of the negative gradient for the multipliers $  \beta^{[k]}_{g} $, $ \nu_{g} $, respectively. According  to equation \eqref{var2}, the Lagrangian multiplier    $   \beta^{[k]}_{g} $ works as a message to ensure that the power allocated to the private message of user $ k $ satisfies the maximum power allowed for each private message. While in equation \eqref{var2}, $ \nu_{g} $ works to ensure that the overall power allocated to the common and private messages intended to $ K_g $ users satisfies the maximum power constraint of group $ g $. Note that at each  iteration, the multipliers $ \beta^{[k]}_{g} $ and $ \nu_{g}  $ are updated until the optimal  values $ P^{*[g]}_{c}  $ and $ P^{*[k,g]}_{p}  $ are found.

\subsubsection{Optimality at $ \lambda_{g} $ and $ \xi_{g}$ values }
Interestingly, the multipliers $ \lambda_{g} $ and $ \xi_{g}$ must also be updated to guarantee high quality of service for the $ K_g $ users  of each group $ g $, where at their optimal values, the power allocated to the private and common messages determined from equation \eqref{grad2} and \eqref{gradcom}, respectively, at the optimal values of $ \beta^{*[k]}_{g} $ and $ \nu^{*}_{g}  $  must be modified to satisfy  the first and second constraints in equation \eqref{OP10}. First, at a given value of $ \xi_{g}$, the optimal value of $ \lambda_{g} $  can be determined by applying the KKT conditions. That is, 
$\frac{\partial \mathcal{F}} {\partial \Gamma } =0$, where $ \Gamma =\min\limits_{g \in \mathcal{G}}  \Big\{ {R^{[g]}_{\text{sum}}} (P^{[g]}_{c},P^{[k,g]}_{p})- \Upsilon  \Big\} $. From equation \eqref{lag}, $ \sum_{g \in \mathcal{G}} \lambda_{g} =1 $. Therefore, 
 the optimum power value $ P^{*[k,g]}_{p} $ allocated to the private message intended to each  user $ k \in \mathcal{K}_{g} $ belonging to group $ g $  can be modified  according to
\begin{multline}
\label{grad22}
\frac{\partial \mathcal{F}_{g} }{\partial P^{[k,g]}_{p} }=  ( {\lambda}_{g}+ \widetilde{\xi}_{g}) \frac{\partial {R^{[k,g]}_{p}} (P^{[k,g]}_{p}) }{\partial P^{[k,g]}_{p} }
- (\delta \lambda_{g}+ \widetilde{\nu}^{*}_{g}+\widetilde{\beta}^{*[k]}_{g}) =0, 
\end{multline}
where $ \widetilde{\xi}_{g}= {\xi}_{g} \sum_{g \in \mathcal{G}} \lambda_{g}$, $ \widetilde{\nu}_{g}={\nu}_{g} \sum_{g \in \mathcal{G}} \lambda_{g}$ and $ \widetilde{\beta}^{[k]}_{g}={\beta}^{[k]}_{g} \sum_{g \in \mathcal{G}} \lambda_{g}$.
Moreover, the optimum power value $ P^{*[g]}_{c} $ allocated to the common message intended to the  $ K_{g} $  users belonging to group $ g $  can be modified  according to

\begin{multline}
\label{gradcom2}
\frac{\partial \mathcal{F}_{g} }{\partial P^{[g]}_{c} }=  ( {\lambda}_{g}+ \widetilde{\xi}_{g}) \frac{\partial {R^{[k,g]}_{c}} ( P^{[g]}_{c}, P^{[k,g]}_{p}) }{\partial  P^{[g]}_{c} }
- (\delta \lambda_{g}+ \widetilde{\nu}^{*}_{g})=0.
\end{multline}
Similar to equations \eqref{var1} and \eqref{var2}, the gradient projection method can be applied to obtain an updated value for the Lagrangian multiplier $ \lambda_{g} $ as follows  
\begin{multline}
\label{var3}
\lambda_{g} (\kappa)= \bigg[\lambda_{g} (\kappa-1)-\epsilon_{3}(\kappa-1) \bigg({R^{[g]}_{\text{sum}}} (P^{[g]}_{c},P^{[k,g]}_{p}) (\kappa-1)-\Upsilon\\ - \min\limits_{g \in \mathcal{G}}  \Big\{ {R^{[g]}_{\text{sum}}} (P^{[g]}_{c},P^{[k,g]}_{p})- \Upsilon  \Big\} \bigg) \bigg]^{+},
\end{multline}
where $ \epsilon_{3}(\kappa-1) $ is a sufficiently small step size at a given iteration $ (\kappa-1)$. Note that, the optimal value $ \lambda^{*}_{g} $ determines the optimal power values allocated to the common and privates messages intended to each $ K_g $ users for a given value $  \xi_{g} \geq 0 $. To find  the optimal value $ \lambda^{*}_{g} $,  according to the first constraint in \eqref{OP10},  $ \Big( {R^{[g]}_{\text{sum}}} (P^{[g]}_{c},P^{[k,g]}_{p})- \Upsilon \Big) \geq \Gamma^{*} $, where $ \Gamma^{*} = \min\limits_{g} \Big\{ {R^{[g]}_{\text{sum}}} (P^{[g]}_{c},P^{[k,g]}_{p})- \Upsilon \Big\} $. 
Let $ \widetilde{G} $ give the total number of the groups with $ \Big( {R^{[g]}_{\text{sum}}} (P^{[g]}_{c},P^{[k,g]}_{p})- \Upsilon \Big)> \Gamma^{*}$, and then, using the KKT conditions, the optimal value $ \lambda^{*}_{g} =0 $ given that 

\begin{equation}
\lambda^{*}_{g} \bigg\{ \Big( {R^{*[g]}_{\text{sum}}} (P^{[g]}_{c},P^{[k,g]}_{p}) - \delta \Big(P^{[g]}_{c}+ \sum\limits_{k\in K_g} P^{*[k,g]}_{p} \Big)\Big)-\Gamma^{*} \bigg\}=0.
\end{equation}
Therefore, for a given value of  $  \xi_{g} \geq 0 $, if $ \Gamma^{*} > \Big({R^{[g]}_{\text{sum}}} (P^{[g]}_{c},P^{[k,g]}_{p})- \Upsilon \Big)$,  $ {R^{[g]}_{\text{sum}}} (P^{[g]}_{c},P^{[k,g]}_{p})- \Upsilon $   cannot be less than $ \min\limits_{g} \Big\{{R^{[g]}_{\text{sum}}} (P^{[g]}_{c},P^{[k,g]}_{p})- \Upsilon\Big\} $.  As a consequence, the optimal value for any group $ g' $, $ g' \neq g $ not within the group set    $ \widetilde{G} $ is obtained when  $ \lambda^{*}_{g} > 0 $  and $ \Gamma^{*} = \Big({R^{[g]}_{\text{sum}}} (P^{[g]}_{c},P^{[k,g]}_{p})- \Upsilon \Big)$. Otherwise, $ \Big( {R^{[g]}_{\text{sum}}} (P^{[g]}_{c},P^{[k,g]}_{p})- \Upsilon \Big)> \Gamma^{*}$ for any group with $ \lambda^{*}_{g}=0 $.  

The optimal value of $  \xi^{*}_{g} $ must also be determined according to  
\begin{multline}
\label{var4}
\xi_{g} (\kappa)= \bigg[\xi_{g} (\kappa-1)-\epsilon_{4}(\kappa-1) \\ \bigg({R^{[g]}_{\text{sum}}} (P^{[g]}_{c},P^{[k,g]}_{p}) (\kappa-1)- R^{[g]}_{min} \bigg) \bigg]^{+},
\end{multline}
where $ \epsilon_{4} $ is a sufficiently small step size. It is worth pointing out that the overall algorithm iterates over the power allocated to the common and private messages intended to each $ K_g $ users until the optimal values $ P^{*[g]}_{c}   $ and $ P^{*[k,g]}_{p}   $ are determined, which can significantly  maximize the sum rate of the $ K_g $ users belonging to each group with respect to power consumption.
Finally,  the computational complexity of  the overall algorithm can be determined  according to the complexity of the dual problem. Therefore, considering the implementation of the gradient method, the computational  complexity is given by $   \mathcal{O}(T \sum^{G}_{g=1} K_{g})$, where $ T $ is the total number of the  iterations. {It is worth mentioning that the traditional BIA, RS or NOMA scheme employs fixed power allocation under the methodology of  MUI management for each scheme.  Moreover, Baselines 1 and 2 are schemes proposed in the next section, where  Baseline 1 employs uniform power allocation among $G$ groups, and Baseline 2 employs the  rate-maximization-based power allocation formulated in this section with a computational  complexity  given by $   \mathcal{O}(T \sum^{L}_{l=1} (K_{l}+K_{eg}))$, where $K_{l}$ are users assigned to each optical AP, and $K_{eg}$ are users located at the edges of the cells.} 
  
\begin{table}
\centering
\caption{Simulation Parameters}
\label{tabla2}
\begin{tabular}{|c|c|}
\hline
Parameter	& Value \\\hline
Transmitter Bandwidth	& 1.5 GHz \\\hline
Spectral response range  & 950-1700 nm \\\hline
Laser wavelength  & 1550 nm \\\hline
Laser beam waist & $ 8 \mu $m \\\hline
Physical area of the photodiode	&15 $\text{mm}^2$ \\\hline
Receiver FOV	& 60 deg \\\hline
Detector responsivity 	& 0.9 A/W \\\hline
Gain of optical filter & 	1.0 \\\hline
Laser noise	& $-155~ dB/H$z \\\hline
\end{tabular}
\end{table}
\section{performance evaluation}
\label{sec:re}
In an indoor environment with  8m$ \times $ 8m$  \times $ 3m dimensions, $ L= 4\times 4 $ optical APs are deployed on the ceiling where each optical AP is an array of $ L_{v} \times L_{v} $ VCSELs. The transmit power of the VCSEL is set to $ 60  $mW \cite{mo6963803}, which is determined according to the so-called divergence angle   $\Theta_{D}= \Theta_{F}/ \sqrt{2 \ln (2) }  $, where $ \Theta_{F}=\frac{\lambda}{\pi W_{0}}  $ is defined  as the angle for full width at half maximum intensity points. On a communication floor located at 2m distance from the ceiling, 
 $ K=20 $ users are distributed randomly, where each user is equipped with an optical detector consisting of $ M $ photodiodes, which has the ability to provide a set of linearly independent channel responses  and  guarantee connectivity to almost all the optical APs in the room \cite{MPGV18,8636954}. All other simulation parameters are listed in \textbf{Table 1}. 
 
Interestingly, two baseline RS schemes are designed taking into account the most common strategies investigated in the literature to test the effectiveness of the proposed BIA-RS scheme and the formulated power allocation problem. The baseline schemes work as follows

\begin{itemize}
\item \textbf{ Baseline 1:} The same strategy of the proposed BIA-RS transmission is applied. However, a uniform power allocation approach is implemented among the messages intended to all the groups to avoid complexity \cite{8636954,8935164}. The power allocated to a certain group is determined by $  P^{[g]}_{max}= \frac{P_{T}}{G} $,  where $ P^{[g]}_{max}= P^{[g]}_{c}+\sum_{k \in \mathcal{K}_{g}} P^{[k,g]}_{p}$. Moreover,  the power allocated to the common message $ P^{[g]}_{c} $  is determined according to \cite{Kh9500371}, while the power allocated to each private message is $ P^{[k,g]}_{p}=\frac{( P^{[g]}_{max}-P^{[g]}_{c})}{K_g} $. It is worth mentioning that Baseline 1 ignores the constraint that ensures high quality of service, resulting in inefficient use of the power budget. 

\item \textbf{ Baseline 2:} The CU has information on the distribution of the users.  Therefore, it can assign each  user located at the center of a cell to the respective optical AP \cite{9064520}, while the users located at the edges of the cells are cooperatively served by all the optical APs considering the implementation of CoMP, detailed mathematical expressions are presented in \cite{6825144}. 
Note that, Baseline 2 cancels ICI at the edge users, while the users assigned to their corresponding optical APs might experience ICI from  the adjacent optical APs. Moreover, MUI within the coverage area of each optical AP is  managed using RS. For power allocation, the optimization problem formulated in Section \ref{sec:power} can be easily rewritten in the context of maximizing the minimum sum rate within the coverage of each optical AP in order to conduct a fair comparison with the proposed BIA-RS scheme.
\end{itemize}   

\begin{figure}[t]
\begin{center}\hspace*{0cm}
\includegraphics[width=1\linewidth]{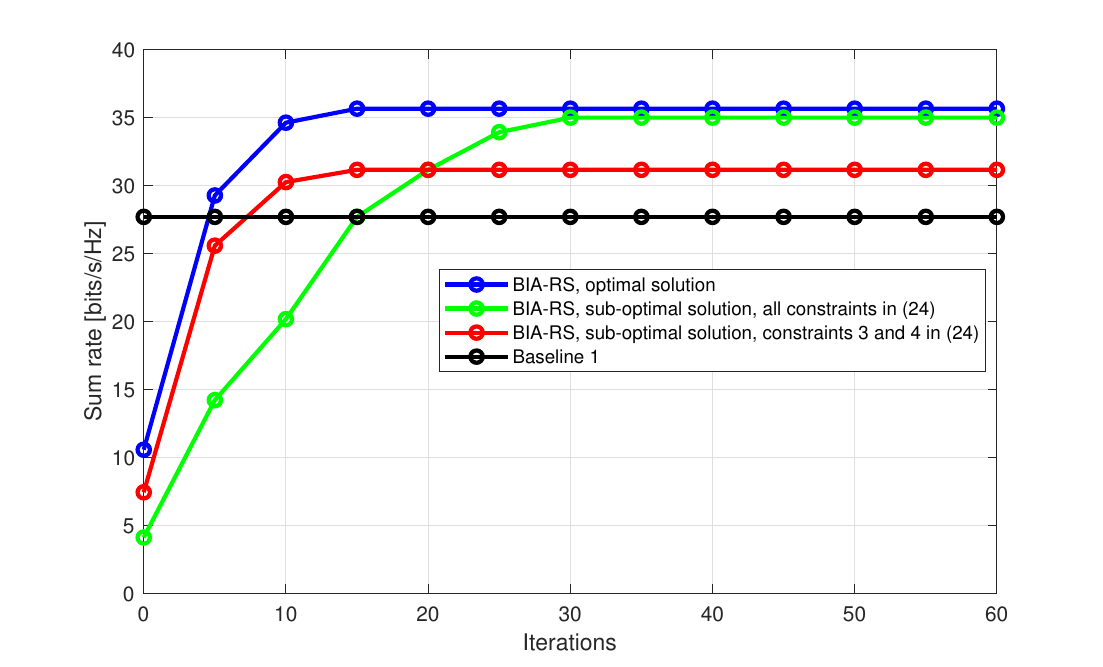}
\end{center}
\vspace{-2mm}
\caption{The sum rate of the proposed BIA-RS scheme after power allocation under different constraints. SNR= 30 dB and $ K=20 $.}\label{RS_fig2}
\vspace{-5mm}
\end{figure}

In Fig.~\ref{RS_fig2}, the performance of the proposed BIA-RS scheme is shown against a set of iterations after performing the optimization problem of power allocation. It can be seen that the sum rate of the proposed scheme can be enhanced considerably after maximizing the minimum sum rate of the users belonging to each group where the power budget is utilized efficiently compared to simply dividing the power among the common and private messages  intended to the users regardless of their requirements as in Baseline 1. The figure further shows that the reformulation of the optimization problem using four Lagrangian multipliers  provides solutions close to the optimal at low complexity. Specifically, the optimization problem is solved using the multipliers $ \beta^{[k]}_{g} $ and $ \nu_{g} $ in the red curve, and therefore, it can be seen that the sum rate is around 31 [bits/s/Hz] compared to slightly above 35 [bits/s/Hz] achieved from solving the main power allocation problem. This is because of allocating the power to maximize the sum rate of each group without guaranteeing  high quality of service for the users, i.e., without the constraints  $ \min\limits_{g \in \mathcal{G}}  \Big\{ {R^{[g]}_{\text{sum}}} (P^{[g]}_{c},P^{[k,g]}_{p})- \Upsilon \Big\} \leq {R^{[g]}_{\text{sum}}} (P^{[g]}_{c},P^{[k,g]}_{p})- \Upsilon
$ and $ {R^{[g]}_{\text{sum}}} (P^{[g]}_{c},P^{[k,g]}_{p}) \geq R^{[g]}_{min} $. On the other hand, when the overall algorithm iterates over the four multipliers $ \beta^{[k]}_{g} $, $ \nu_{g} $, $ \lambda_{g} $ and $ \xi_{g}$ in the green curve, a significant solution is provided close to the optimal one. This solution is considered for the rest of the results due  to its practicality in terms of complexity.

\begin{figure}[t]
\begin{center}\hspace*{0cm}
\includegraphics[width=1\linewidth]{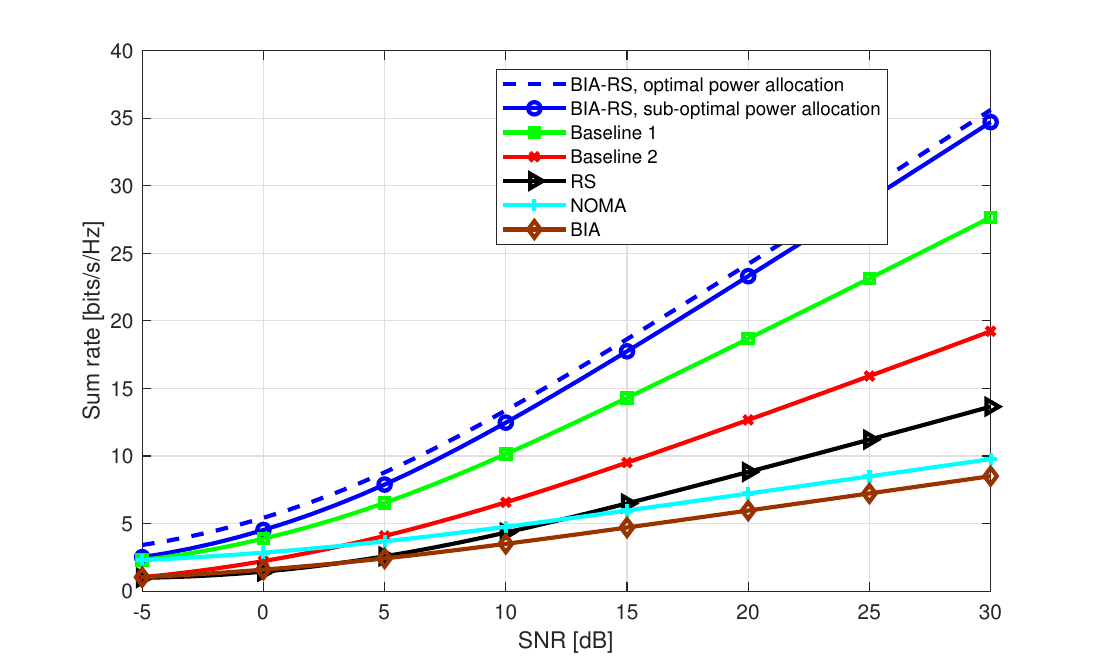}
\end{center}
\vspace{-2mm}
\caption{The sum rate of the proposed BIA-RS scheme versus SNR compared with benchmark schemes. $ K=20 $.}\label{RS_fig3_2}
\vspace{-5mm}
\end{figure}

The sum rate of the proposed BIA-RS scheme is shown in Fig.~\ref{RS_fig3_2} against a range of SNR values to demonstrate its superiority compared to benchmarking schemes that work with different interference management strategies. {It is shown that the proposed BIA-RS scheme (sub-optimal solution) provides sum rates close to the optimal solution for $T$ beyond 30 (see Fig.~\ref{RS_fig2}), and higher than other transmission schemes at different SNR values}. Compared to Baselines 1 and 2, the proposed BIA-RS scheme coordinates the transmission among all the optical APs determining the outer precoding matrices of the groups, and it allocates the power budget to ensure high quality of service at low power consumption.  The application of Baseline 1  allocates fixed  power to each message without considering rate-maximization, while in Baseline 2, users assigned to optical APs  based on their locations experience  high ICI from the neighboring optical APs. 
Among RS, NOMA and BIA, RS implemented within the coverage area of each optical AP  provides high sum rates as the SNR increases due to allocate a fraction of the power to the common message \cite{Kh9500371} that gives the users the opportunity to cancel MUI, however, the performance of traditional RS is subject to ICI and significant  noise enhancement. In BIA \cite{8636954}, all the optical APs cooperatively serve the users, i.e., the number of optical APs and users determines the length of the transmission block, and each user must subtract the information transmitted to $ K-1 $ users. Therefore, the sum rate of the users using BIA slightly increases with the increase of SNR. Moreover, the low performance of NOMA proposed in \cite{7342274} within the coverage area of each optical AP  is due to the comparable channel gains of the users, i.e., high residual noise resulting from SIC,  and the high interference received from the adjacent APs. From now on, we focus our attention on RS-based transmission to understand the proposed schemes.

\begin{figure}[t]
\begin{center}\hspace*{0cm}
\includegraphics[width=1\linewidth]{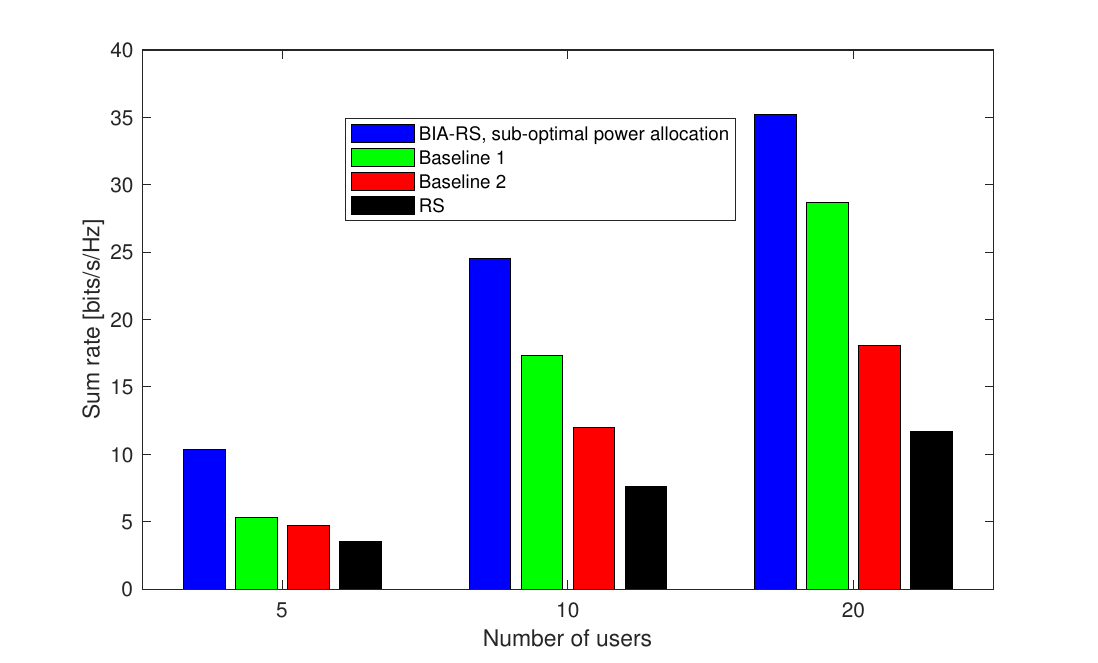}
\end{center}
\vspace{-2mm}
\caption{The sum rate of the proposed BIA-RS scheme versus $ K $ users compared with benchmark schemes. }\label{RS_fig5}
\vspace{-5mm}
\end{figure}

In Fig.~\ref{RS_fig5}, the performance of the proposed BIA-RS scheme is depicted against different numbers of users  compared to the counterpart schemes. It can been seen that the sum rate of the network increases with the number of users regardless of the  transmission scheme implemented for interference management since it represents the aggregate data rate determined from the sum of the user data rates. The proposed BIA-RS scheme is more suitable for the OWC network as the number of users increases, providing higher sum rates in all the scenarios considered  due to the fact that full coordination among the optical APs is achieved, where the users receive useful information from the neighboring optical APs. It is worth mentioning that the overall power allocation algorithm  further enhances the performance of the proposed BIA-RS scheme compared to the fixed power allocation method implemented in Baseline 1. Despite the implementation of the power allocation algorithm in Baseline 2, each user receives low data rate limited because of relatively high ICI received at the users assigned to each optical APs considering the characteristics of the transmitters used. The traditional RS proposed in \cite{Kh9500371} shows low performance compared to the other schemes due to the use of fixed power allocation among the messages sent by each optical APs, as well as the high ICI generated among the optical APs.

\begin{figure}[t]
\begin{center}\hspace*{0cm}
\includegraphics[width=1\linewidth]{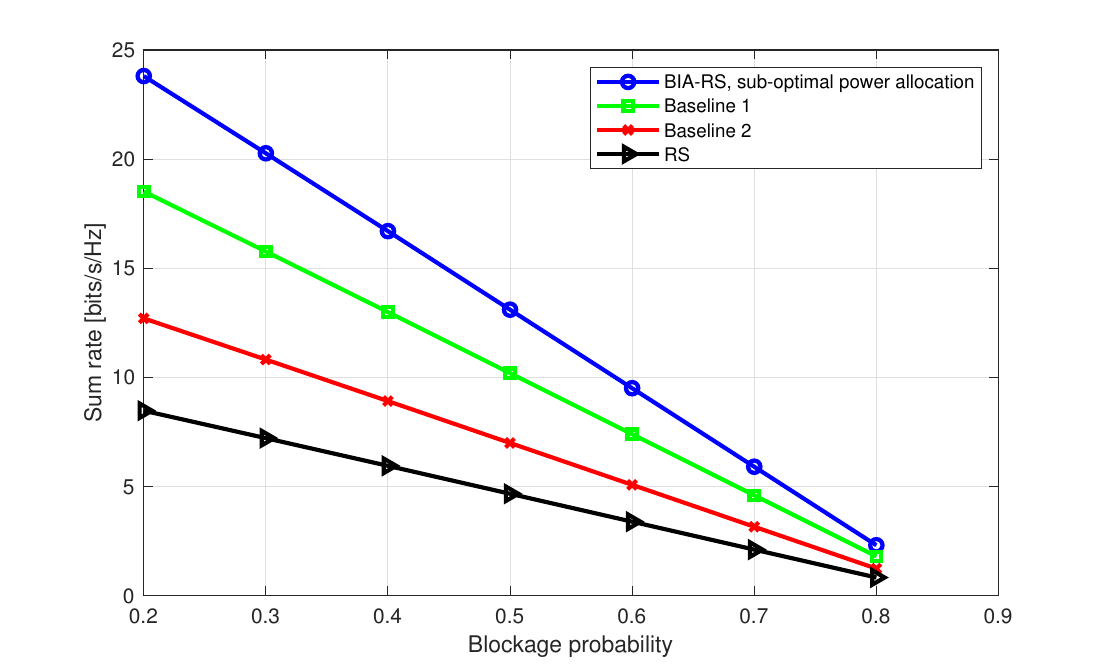}
\end{center}
\vspace{-2mm}
\caption{Sum rates against blockage probability. $ K=20 $.}\label{RS_fig7}
\vspace{-5mm}
\end{figure} 
Fig.\ref{RS_fig7} shows the sum rate of the network versus blockage probability. The blockage of the LoS link in OWC can be easily caused by any object in the room.  It is shown that the sum rate decreases  considerably as the blockage probability increases where each user receives low channel gain. However, the proposed BIA-RS scheme is robust compared to the other schemes considered due to the fact that  each user receives useful information from the whole set of the optical APs, and that the power is allocated to maximize the minimum sum rate of each group. Note that, if a certain user receives low channel gain, more power might be allocated to meet the requirements of the minimum sum rate. In our network, blocked users might experience handover to a neighboring optical AP or to another system in the room, i.e., WiFi, depending on the transmission scheme implemented. However, it is beyond the scope of this paper. 
\begin{figure}[t]
\begin{center}\hspace*{0cm}
\includegraphics[width=1\linewidth]{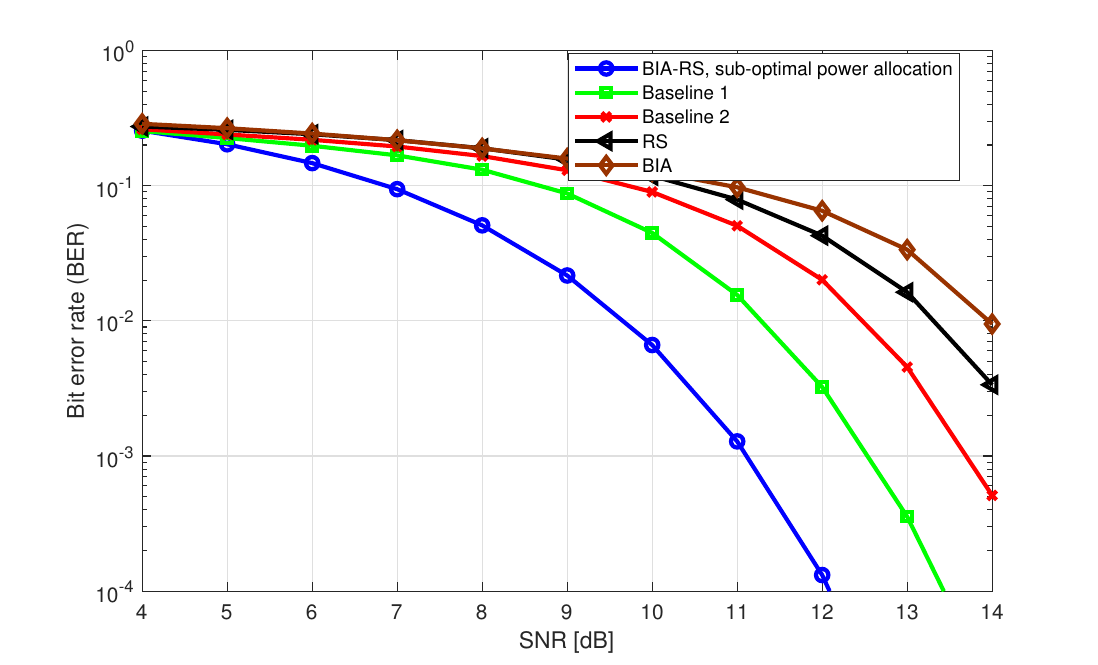}
\end{center}
\vspace{-2mm}
\caption{BER for 2-PAM Modulation. $ K=20 $.}\label{RS_fig4}
\vspace{-5mm}
\end{figure} 

To evaluate the BER of the proposed BIA-RS scheme, multi-level N-ary pulse amplitude modulation (N-PAM) is considered  where the mean of the transmitted signal is equal to the current, which ensures the linear response of the optical transmitter. In Fig.~\ref{RS_fig4}, BER is plotted against a range of  SNR values, (4-14) dB, considering  different transmission schemes. It can be seen that the proposed BIA-RS is superior compared to Baselines 1 and 2, RS and BIA at different SNR values. Interestingly, the implementation of BIA to coordinate the transmission among the optical APs achieves a BER value less than $ 10^{-4} $ at 13 dB  and 14 dB SNRs with and without performing the  power allocation algorithm, respectively. On the other hand, Baseline 2, RS and BIA  achieve BER values less than  $ 10^{-3} $, less than  $ 10^{-2} $ and  $ 10^{-2} $,  respectively, at 14 dB SNR. This is due to the fact that Baseline 2 does not cancel ICI for all the users, and  all the  users in RS experience high ICI from the neighboring cells, in addition to the fixed power allocation approach among the common and private messages. Moreover, BIA suffers noise enhancement due to the high number of users.

\begin{figure}[t]
\begin{center}\hspace*{0cm}
\includegraphics[width=1\linewidth]{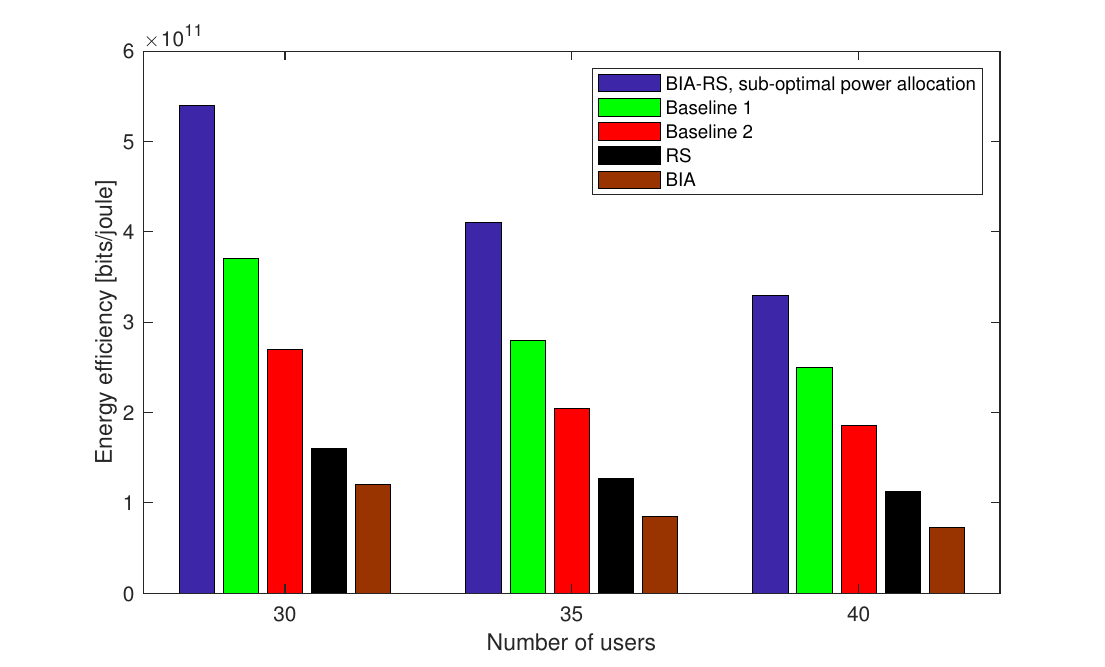}
\end{center}
\vspace{-2mm}
\caption{Energy efficiency of the proposed BIA-RS scheme versus $ K $ users.}\label{RS_fig6}
\vspace{-5mm}
\end{figure}

To further test the performance of the proposed BIA-RS scheme, the energy efficiency of the network, which is one of the most important metrics in wireless communications, is depicted in Fig.~\ref{RS_fig6} against an extremely high number of users $ \{30, 35, 40\} $. It can be seen that the energy efficiency of the network decreases as the number of users increases up to $ 40 $ users due to the fact that  more power is consumed to support such a high number of users. Interestingly, the proposed BIA-RS scheme is more energy efficient  than  the benchmarking schemes for all the scenarios considered. It is expected as  the strategy of the proposed BIA-RS scheme  is more suitable for dense OWC networks, and the power allocation algorithm  distributes the power efficiently among the messages intended to all the users resulting in high aggregate data rates as proven in the previous results. Baseline 1 has higher energy efficiency compared to Baseline 2, RS and BIA since it uses the same methodology as the proposed BIA-RS scheme in coordinating the transmission and managing the interference. 
 
\section{Conclusions}
\label{sec:con}
In this work, a novel RS-based transmission scheme is proposed in an OWC network to serve multiple users simultaneously, maximizing the spectral efficiency of  the network. We first derive the BIA-RS scheme to  achieve three essential points: {\it i)} full coordination among the optical APs,  {\it ii)}  allocate non-orthogonal resource to multiple groups of users with guarantee of inter-group interference management, {\it iii)} enabling parallel RS schemes to work simultaneously in the groups to manage  MUI. After that, a power allocation optimization problem is formulated to maximize the minimum sum rate within each group through finding the optimum power values allocated to the groups  and to the common and private  messages intended to the users belonging to each group. The optimization problem  is reformulated via Lagrangian multipliers to obtain sub-optimal solutions at low complexity. The results show that the proposed BIA-RS scheme achieves high performance in terms of sum rate, BER and energy efficiency compared to BIA, NOMA, RS and two baselines schemes derived according to the majority of the RS work in the literature. 

For future directions, the application of BIA-RS in a very dense OWC network might be considered at the cost of high complexity to determine the precoding matrices of the groups from the whole set of optical APs, and to manage the resources of the network. Therefore, AP-user combinations can be determined to form multiple amorphous intelligent optical cells with soft edges, i.e., no ICI. The strategy of BIA-RS can be applied within each amorphous cell instead of blindly forming large-scale single optical cell in the network. Such scenarios lead to the formulation of optimization problems with different contexts including load balancing, handover, resource management, etc. 

\appendix 
We denote the the maximum permissible exposure as $  E_{e,\max}(t_e) $, where $  t_e  $ is the exposure duration. In \cite{mo6963803}, $  E_{e,\max}(t_e) $ is determined  for different wavelengths. At this point, the level of exposure  to a certain  wavelength at the most hazard position can be determined by  
\begin{multline}
E_{e}(d_h)= \frac{1}{\pi (\frac{d_c}{2})^{2}}
\int_{0}^{d_c/2 } I(r,d_h) 2\pi r dr \\ =\frac{1}{\pi (\frac{d_c}{2})^{2}} \int_{0}^{d_c/2 } \frac{2 P_{tr}}{\pi W^{2}({d_h})}~ \mathrm{exp}\left(-\frac{2 r^{2}}{W^{2}({d_h})}\right)2\pi r dr\\=
\frac{P_{tr}}{\pi (\frac{d_c}{2})^{2}} \bigg(1-\mathrm{exp}\left(-\frac{d_{c}^{2}}{2 W^{2}({d_h})}\right)\bigg), 
\end{multline}
where $ d_h $ is the hazard distance and   $ d_c  $ is the aperture diameter of the cornea. Interestingly, if  $ E_{e}(d_h)\leq E_{e,\max}(t_e) $, it means that the transmit power of the VCSEL is safe regardless of the position of the user, otherwise, $ E_{e}(d_h)\geq E_{e,\max}(t_e) $, causing damage to the human eye. Therefore, the transmit power of the VCSEL $ P_{tr} $ must be less or equal to the maximum permissible transmit power determined according to 
\begin{equation}
  P_{\max}= \frac{\pi}{4} d_{c}^{2} E_{e,\max}(t_e) \bigg( 1- \mathrm{exp} \bigg( \frac{d_{c}^{2}}{2 W^{2}({d_h})}\bigg) \bigg)^{-1}.
\end{equation}

\bibliographystyle{IEEEtran}
\bibliography{IEEEabrv,mybib}

\end{document}